\documentclass[preprintnumbers, prd, showpacs, floatfix,twocolumn,
preprintnumbers, letterpaper,
superscriptaddress,nofootinbib]{revtex4}
\usepackage{amsfonts}
\usepackage{amsmath}
\usepackage{amssymb,epsf}
\usepackage{latexsym}
\usepackage{graphicx,epsfig}
\usepackage{amssymb}
\usepackage{subfigure}
\usepackage{epstopdf}
\usepackage[colorlinks=true,citecolor=blue,linkcolor=blue,urlcolor=black]{hyperref}

\begin{document}

\title{Thermodynamics of topological nonlinear charged Lifshitz black holes}
\author{M. Kord Zangeneh}
\email{mkzangeneh@shirazu.ac.ir}
\affiliation{Physics Department and Biruni Observatory, Shiraz University, Shiraz 71454,
Iran}
\author{A. Sheykhi}
\email{asheykhi@shirazu.ac.ir}
\author{M. H. Dehghani}
\email{mhd@shirazu.ac.ir}
\affiliation{Physics Department and Biruni Observatory, Shiraz University, Shiraz 71454,
Iran}
\affiliation{Research Institute for Astronomy and Astrophysics of Maragha (RIAAM), P.O.
Box 55134-441, Maragha, Iran}

\begin{abstract}
In this paper, we construct a new class of analytic topological Lifshitz
black holes with constant curvature horizon in the presence of power-law
Maxwell field in four and higher dimensions. We find that in order to obtain
these exact Lifshitz solutions, we need a dilaton and at least three
electromagnetic fields. Interestingly enough, we find that the reality of
the charge of the electromagnetic field which is needed for having solutions
with curved horizon rules out black holes with hyperbolic horizon. Next, we
study the thermodynamics of these nonlinear charged Lifshitz black holes
with spherical and flat horizons by calculating all the conserved and
thermodynamic quantities of the solutions. Furthermore, we obtain a
generalized Smarr formula and show that the first law of thermodynamics is
satisfied. We also perform a stability analysis in both canonical and
grand-canonical ensemble. We find that the solutions are thermally stable in
a proper ranges of the metric parameters. Finally, we comment on the
dynamical stability of the obtained solutions under perturbations in four
dimensions.
\end{abstract}

\maketitle

\address{Physics Department and Biruni Observatory, College of
Sciences, Shiraz University, Shiraz 71454, Iran} 
\address{ Research Institute for Astronomy and Astrophysics of Maragha
         (RIAAM), P.O. Box 55134-441, Maragha, Iran}

\section{INTRODUCTION}

The idea of correspondence between gravity in an anti-de Sitter spacetime
and the conformal field theory living on its boundary (AdS/CFT
correspondence), was first suggested by Maldacena in $1998$ \cite{Mald}.
According to Maldacena's arguments the effects of the string or M-theory in
a $d$-dimensional AdS$_{n+1}\times S^{d-n-1}$ spacetime can be appeared in
the form of a field theory on an $n$-dimensional $r$-constant brane which is
the boundary of AdS$_{n+1}$ spacetime. He expressed that the finite
temperature configurations in the field theory which lies on the $n$%
-dimensional brane and is decoupled from the bulk correspond to black hole
configurations in AdS$_{n+1}$ spacetime \cite{Mald}. This proposal makes us
able to study the non-perturbative aspects of the field theories. This idea
attracted a lot of interests rapidly and has been studied from different
points of view \cite{AdS}. Since the line element of the AdS spacetime,%
\begin{equation}
ds^{2}=-\frac{r^{2}}{l^{2}}dt^{2}+\frac{l^{2}}{r^{2}}dr^{2}+r^{2}\sum%
\limits_{i=1}^{n-1}dx_{i}^{2},
\end{equation}%
is invariant under an isotropic conformal transformation 
\begin{equation}
t\rightarrow \lambda t,\text{ \ \ \ \ }x_{i}\rightarrow \lambda x_{i},\text{
\ \ \ \ }r\rightarrow \lambda ^{-1}r,
\end{equation}%
so the application of AdS/CFT is restricted to systems respected isotropic
scale invariance. But, quantum critical systems show scaling symmetry

\begin{equation}
t\rightarrow \lambda ^{z}t,\text{ \ \ \ \ }x_{i}\rightarrow \lambda x_{i},
\end{equation}%
where $z>1$ is a dynamical critical exponent which shows the degrees of
anisotropy between space and time. Therefore, we can generalize AdS
spacetime to other spacetimes with anisotropic scale invariance 
\begin{equation}
ds^{2}=-\frac{r^{2z}}{l^{2z}}dt^{2}+{\frac{l^{2}dr^{2}}{r^{2}}}%
+r^{2}\sum\limits_{i=1}^{n-1}dx_{i}^{2}.
\end{equation}%
This spacetime was first introduced in \cite{Lif} and is called Lifshitz
spacetime. Black hole configurations in the Lifshitz spacetime are dual to
the field theories enjoying anisotropic scale invariance on the boundary
(brane).

It is well-known that Lifshitz spacetime is not a vacuum solution of
Einstein gravity and it needs some matter sources. In many works this matter
source was considered to be a massive gauge field \cite%
{Mann,Fixz2,Fixz,NFixz}. Some of these solutions were obtained for fixed $z$ 
\cite{Mann,Fixz2,Fixz}. For instance, the authors of \cite{Mann,Fixz2}
focused on $z=2$ case. The main disadvantage of such models containing
massive gauge fields is that it is usually impossible to obtain an analytic
solution for an arbitrary $z$, although some efforts have been made to do
that \cite{beato}. In Ref. \cite{tarrio}, it was shown that by considering a
dilaton field, instead of a massive gauge field, we are able to derive exact
solutions which helps us to get more insight. In addition, string theory in
its low energy limit recovers Einstein gravity with a scalar dilaton field
nonminimally coupled to gravity and other fields such as gauge fields \cite%
{dilaton}. Thermal behavior of uncharged Lifshitz black branes in
Einstein-dilaton gravity has been investigated in \cite{peet}.
Thermodynamics of linearly charged Lifshitz black hole/branes in the context
of Einstein-dilaton gravity has been studied in \cite{tarrio}.

Matter sources possessing conformal invariance have been always of much
interests \cite{conf}. The first black hole solution with a conformally
invariance matter source is the Reissner-Nordstr\"{o}m solution in four
dimensions. In higher dimensions, Maxwell source is no longer conformally
invariant. However, as in the case of massless Klein-Gordon Lagrangian, a
particular power of Maxwell Lagrangian can respect conformal invariance in
arbitrary dimensions \cite{hassaine}. In fact, in ($n+1$)-dimensional
spacetime the action

\begin{equation}
S_{\text{PM}}=\int_{\mathcal{M}}d^{n+1}x\sqrt{-g}\mathcal{L}_{PM},
\end{equation}%
where

\begin{equation*}
\mathcal{L}_{\text{PM}}=\left[ -e^{-4/(n-1)\lambda \Phi }F_{\mu \nu }F^{\mu
\nu }\right] ^{(n+1)/4}
\end{equation*}%
is conformally invariant, i.e., it is invariant under the conformal
transformation $g_{\mu \nu }\rightarrow \Omega ^{2}g_{\mu \nu }$, $A_{\mu
}\rightarrow A_{\mu }$ and $\Phi \rightarrow \Phi $. Black hole solutions
with power-law Maxwell electrodynamic fields have been studied in \cite%
{hassaine,hasmart,confel,MHM,confel2}.

Here, we shall consider a class of asymptotically Lifshitz black hole
solutions of Einstein-dilaton gravity with power-law Maxwell field. Although
electrodynamic Lagrangian is conformally invariant in $(n+1)/4$ dimensions,
this does not implies that the power should be fixed for general cases. We
first consider a four dimensional toy model and fully describe the procedure
of solving field equations and gaining desired solutions. Then, we extend
our solutions to the higher-dimensional spacetimes. From the prediction of
string theory, we know that the spacetime may have more than four
dimensions. Therefore we get enough motivation for studying black hole
solutions of Einstein gravity in all higher dimensions. Although it was
thought for a while that the extra spacial dimensions are of the order of
the Planck scale, recent theories suggest that if we live on a three
dimensional brane embedded in a higher dimensional bulk, it is quite
possible to have the extra dimensions relatively large and still
unobservable \cite{high,high2}. In this scenario, all gravitational objects
including black holes are higher-dimensional.

This paper is structured as follows. In the next section, we introduce the
four dimensional action in the presence of a general nonlinear
electrodynamic and derive the equations of motion by varying the action. We
also give some general remarks about nonlinearly charged topological
Lifshitz black holes. Furthermore, focusing on power-law nonlinear
electrodynamic source, we comprehensively describe the procedure of
obtaining asymptotic Lifshitz topological black hole exact solutions. In
section \ref{ndim}, we generalize our study to higher dimensions. In section %
\ref{therm}, we calculate the mass of black hole solutions by the modified
Brown-York method and study thermodynamics of the obtained solutions. We
also check the validity of the first law of thermodynamics by finding the
Smarr formula for the mass. In section \ref{stab}, we investigate thermal
stability of the solutions in both canonical and grand-canonical ensembles.
The dynamical stability of $4$-dimensional AdS black holes is examined in %
\ref{dystab}. In the last section we give a summary and closing remarks on
our work.

\section{ACTION AND ASYMPTOTIC LIFSHITZ SOLUTIONS\label{4dim}}

The action of Einstein-dilaton gravity in the presence of a nonlinear
electromagnetic and two linear Maxwell fields can be written as:\textbf{\ }%
\begin{eqnarray}
S &=&-\frac{1}{16\pi }\int_{\mathcal{M}}d^{4}x\sqrt{-g}\left\{ \mathcal{R}%
-2(\nabla \Phi )^{2}\right.  \notag \\
&&\left. -2\Lambda _{4}+\mathcal{L}(F,\Phi
)-\sum\limits_{i=2}^{3}e^{-2\lambda _{i}\Phi }H_{i}\right\} ,  \label{action}
\end{eqnarray}%
where\textbf{\ }$\mathcal{R}$\textbf{\ }is the Ricci scalar on manifold%
\textbf{\ }$\mathcal{M}$\textbf{, }$\Phi $\textbf{\ }is the dilaton field, $%
\Lambda _{4}$\ and $\lambda _{i}$'s are constant parameters. In Eq. (\ref%
{action}) $F$\ and $H_{i}$'s are the Maxwell invariants of electromagnetic
fields $F_{\mu \nu }=\partial _{\lbrack \mu }A_{\nu ]}$\ and $\left(
H_{i}\right) _{\mu \nu }=\partial _{\lbrack \mu }\left( B_{i}\right) _{\nu
]} $, where $A_{\mu }$\ and $\left( B_{i}\right) _{\mu }$s are the
electromagnetic potentials.\textbf{\ }$\mathcal{L}(F,\Phi )$\textbf{\ }%
stands for nonlinear electromagnetic Lagrangian. Varying the action (\ref%
{action}) with respect to the metric $g_{\mu \nu }$, the dilaton field $\Phi 
$, electromagnetic potentials $A_{\mu }$\ and $\left( B_{i}\right) _{\mu }$%
's, lead to the following field equations%
\begin{eqnarray}
&&\mathcal{R}_{\mu \nu }=\frac{1}{2}g_{\mu \nu }\left[ 2\Lambda _{4}+2%
\mathcal{L}_{F}F-\mathcal{L}(F,\Phi )\right.  \notag \\
&&\left. -\sum\limits_{i=2}^{3}H_{i}e^{-2\lambda _{i}\Phi }\right]
+2\sum\limits_{i=2}^{3}e^{-2\lambda _{i}\Phi }\left( H_{i}\right) _{\mu
\lambda }\left( H_{i}\right) _{\nu }^{\text{ \ }\lambda }  \notag \\
&&-2\mathcal{L}_{F}F_{\mu \lambda }F_{\nu }^{\text{ \ }\lambda }+2\partial
_{\mu }\Phi \partial _{\nu }\Phi ,  \label{FE1}
\end{eqnarray}%
\begin{eqnarray}
&&\nabla ^{2}\Phi +\frac{\mathcal{L}_{\Phi }}{4}+\sum\limits_{i=2}^{3}\frac{%
\lambda _{i}}{2}e^{-{2\lambda _{i}\Phi }}H_{i}=0,  \label{FE2} \\
&&\triangledown _{\mu }\left( \mathcal{L}_{F}F^{\mu \nu }\right) =0,
\label{FE3} \\
&&\nabla _{\mu }\left( e^{-{2\lambda _{i}\Phi }}\left( H_{i}\right) ^{\mu
\nu }\right) =0.  \label{FE4}
\end{eqnarray}%
where we use the convention $X_{Y}=\partial X/\partial Y$. Our purpose here
is to find asymptotic Lifshitz topological black hole solutions of the field
equations (\ref{FE1})-(\ref{FE4}). The line elements of such a metric in
four dimensions can be written as \cite{Mann,tarrio} 
\begin{equation}
ds^{2}=-\frac{r^{2z}}{l^{2z}}f(r)dt^{2}+{\frac{l^{2}}{r^{2}}}\frac{dr^{2}}{%
f(r)}+r^{2}d\Omega _{k}^{2},  \label{met}
\end{equation}%
where

\begin{equation}
d\Omega _{k}^{2}=\left\{ 
\begin{tabular}{ll}
$d\theta ^{2}+\sin ^{2}(\theta )d\phi ^{2}$\ \ \ \ \  & $k=1$ \\ 
$d\theta ^{2}+d\phi ^{2}$ & $k=0$ \\ 
$d\theta ^{2}+\sinh ^{2}(\theta )d\phi ^{2}$ & $k=-1$%
\end{tabular}%
\right. ,
\end{equation}%
represents a $2$-dimensional hypersurface with constant curvature $2k$\ and
volume $\omega $. For $k=1$, the topology of the event horizon is a
two-sphere $S^{2}$, and the spacetime has the topology $R^{2}\times S^{2}$.
For $k=0$, the topology of the event horizon is that of a torus and the
spacetime has the topology $R^{2}\times T^{2}$. For $k=-1$, the surface $%
\Sigma $\ is a $2$-dimensional hypersurface $H^{2}$\ with constant negative
curvature. In this case the topology of spacetime is $R^{2}\times H^{2}$. It
is worth mentioning that the asymptotic behavior of metric (\ref{met})
should be%
\begin{equation}
f(r)=1+\frac{kl^{2}}{z^{2}r^{2}}.  \label{Asym}
\end{equation}%
Note that this asymptotic behavior is the exact Lifshitz solution which is
obtained in Ref. \cite{Mann} for $z=2$\ and also it reduces to AdS solution
for $z=1$\ as one expects.

In this paper, we intend to consider the power-law Maxwell nonlinear source,
but here we pause to present some general results coming from general
nonlinear electromagnetic sources. Using (\ref{met}), (\ref{FE4}) can
immediately be integrated as%
\begin{equation}
\left( H_{i}\right) _{rt}=q_{i}r^{z-3}e^{2\lambda _{i}\Phi }.  \label{HH}
\end{equation}%
Substituting the line element (\ref{met}) and (\ref{HH}) to (\ref{FE2}), one
obtains

\begin{equation}
4\left( {r^{3+z}f\Phi ^{\prime }}\right) ^{\prime }=-{{l}^{2}r^{1+z}}%
\mathcal{L}{_{\Phi }+}\frac{4{l^{2\,z}}}{{r}^{3-z}}{\sum\limits_{i=2}^{3}%
\lambda _{i}{q}_{i}^{2}e{^{2\lambda _{i}\Phi },}}  \label{E1}
\end{equation}%
where the prime denotes derivative with respect to $r$. Let's consider the
case where $\lambda _{i}$'s are positive (later, we see that this is true)
and $\mathcal{L}_{\Phi }<0$\ (in most of nonlinear electromagnetic cases
this is a true consideration \cite{stef}). In this case the RHS of (\ref{E1}%
) is always positive and therefore $r^{3+z}f\Phi ^{\prime }$\ is an
increasing function. On the other hand, we can prove that our black hole has
just one horizon in this case. Consider that there are two inner and outer
horizons (denoted by $r_{-}$\ and $r_{+}$\ respectively). In one side $%
r^{3+z}f\Phi ^{\prime }$ is an increasing function, while on the other side $%
4r^{3+z}\Phi ^{\prime }f\mid _{r_{+}}-4r^{3+z}\Phi ^{\prime }f\mid
_{r_{-}}=0 $. This contradiction means that our assumption of having inner
and outer horizon is not correct. Thus, if a black hole exists, it has just
one horizon and its causal structure is Schwarzschild-like. Since an
increasing function has just one zero, if there is any, the zero of $%
r^{3+z}f\Phi ^{\prime }$\ occurs at the horizon when we have black hole
solutions. Consequently, $r^{3+z}f\Phi ^{\prime }<0$\ inside the black hole
horizon and $r^{3+z}f\Phi ^{\prime }>0$\ outside of it and therefore since $%
f<0$\ and $f>0$, inside and outside the horizon of black hole respectively ,
we always have $\Phi ^{\prime }>0$.

After presenting the above general results about a general nonlinear
electromagnetic sources, we continue our study with power-law Maxwell source
with the Lagrangian $\mathcal{L}(F,{\Phi })=\left( -e^{-2\lambda _{1}\Phi
}F\right) ^{p}$. In this case, one can integrate the electromagnetic field
equation (\ref{FE3}) as%
\begin{equation}
F_{rt}=q_{1}r^{[(2p-1)z-2p-1]/(2p-1)}e^{(2\lambda _{1}p\Phi )/(2p-1)}.
\label{FF}
\end{equation}%
Substituting (\ref{met}), (\ref{HH}) and (\ref{FF}) into Eqs. (\ref{FE1})
and (\ref{FE2}), one receives 
\begin{widetext}
\begin{eqnarray}
&&{\frac{rf^{\prime }+3\,f+{r}^{2}\Phi ^{\prime 2}f}{{l}^{2}}+}\Lambda _{4}-%
\frac{k}{r^{2}}+\frac{\left( 2\,p-1\right) q_{1}^{2\,p}e{^{{(2\lambda
_{1}p\Phi )/(2\,p-1)}}}}{{2}^{1-p}{l}^{2\,\left( 1-z\right) p}{r}^{{%
4p/(2\,p-1)}}}+\sum\limits_{i=2}^{3}{\frac{e{^{2\lambda _{i}\Phi }q}_{i}^{2}%
}{{l}^{2-2\,z}{r}^{4}}}=0,  \label{fe1} \\
&&{\frac{rf^{\prime }+f+2\,fz-{r}^{2}\Phi ^{\prime 2}f}{{l}^{2}}+}\Lambda
_{4}-\frac{k}{r^{2}}+\frac{\left( 2\,p-1\right) q_{1}^{2\,p}e{^{{(2\lambda
_{1}p\Phi )/(2\,p-1)}}}}{{2}^{1-p}{l}^{2\,\left( 1-z\right) p}{r}^{{%
4p/(2\,p-1)}}}+\sum\limits_{i=2}^{3}{\frac{e{^{2\lambda _{i}\Phi }q}_{i}^{2}%
}{{l}^{2-2\,z}{r}^{4}}}=0,  \label{fe2} \\
&&{\frac{{r}^{2}{f}^{\prime \prime }+3\,r\left( 1+z\right) {f}^{\prime }+2\,{%
r}^{2}\Phi ^{\prime 2}f+2\left( \,{z}^{2}+\,z+1\right) f}{2{l}^{2}}+}\Lambda
_{4}-\frac{q_{1}^{2\,p}e{^{{(2\lambda _{1}p\Phi )/(2\,p-1)}}}}{{2}^{1-p}{l}%
^{2\,\left( 1-z\right) p}{r}^{{4p/(2\,p-1)}}}-\sum\limits_{i=2}^{3}{\frac{e{%
^{2\lambda _{i}\Phi }q}_{i}^{2}}{{l}^{2-2\,z}{r}^{4}}}=0,  \label{fe3} \\
&&{\frac{r^{2}f\Phi ^{\prime \prime }+\left( r^{2}f^{\prime }+\left(
3+z\right) rf\right) \Phi ^{\prime }}{{l}^{2}}-}\frac{{\lambda _{1}p}%
q_{1}^{2\,p}e{^{{(2\lambda _{1}p\Phi )/(2\,p-1)}}}}{{2}^{1-p}{l}^{2\,\left(
1-z\right) p}{r}^{{4p/(2\,p-1)}}}{-\sum\limits_{i=2}^{3}\frac{\lambda _{i}{q}%
_{i}^{2}e{^{2\lambda _{i}\Phi }}}{{l}^{2-2\,z}{r}^{4}}}=0.  \label{fe4}
\end{eqnarray}
\end{widetext}Subtracting (\ref{fe2}) from (\ref{fe1}), we arrive at 
\begin{equation}
{(1-z)+{r}^{2}\Phi ^{\prime 2}=0},
\end{equation}%
with the solution 
\begin{equation}
\Phi (r)=\sqrt{z-1}\ln \left( \frac{r}{b}\right) ,  \label{Phi}
\end{equation}%
where $b$ is a constant and $z\geq 1$. With (\ref{Phi}) in hand, by solving
field equations (\ref{fe2})-(\ref{fe4}), we can find metric function as%
\begin{eqnarray}
&&f(r)=\frac{kl^{2}}{zr^{2}}-\frac{\Lambda _{4}l^{2}}{z+2}-\frac{m}{r^{z+2}}
\notag \\
&&-\frac{{2}^{p-1}q_{1}^{2\,p}{l}^{2\,p(z-1)+2}\left( 2p-1\right) ^{2}{b}%
^{\left( -2\,{\sqrt{-1+z}{\lambda _{1}}p}\right) /\left( 2\,p-1\right) }}{%
\left( 2\sqrt{-1+z}{\lambda _{1}}\,p+z(2p-1)-2\right) {r}^{2p\left(
2-\lambda _{1}\sqrt{z-1}\right) /\left( 2p-1\right) }}  \notag \\
&&-\sum\limits_{i=2}^{3}{\frac{q_{i}^{2}{l}^{2\,z}{r}^{-4+2\,\sqrt{-1+z}%
\lambda _{i}}}{\left( -2+z+2\,\sqrt{-1+z}\lambda _{i}\right) {b}^{2\,\sqrt{%
-1+z}\lambda _{i}}}},
\end{eqnarray}%
The field equations are fully satisfied provided: 
\begin{eqnarray}
q_{2}^{2} &=&\frac{-\Lambda _{4}\left( z-1\right) b^{4}}{\left( z+1\right)
l^{2(z-1)}},  \notag \\
q_{3}^{2} &=&\frac{kb^{2}\left( z-1\right) }{l^{2(z-1)}z},  \notag \\
\lambda _{1} &=&\frac{\left( 1-2p\right) \sqrt{z-1}}{p},  \notag \\
\lambda _{2} &=&\frac{2}{\sqrt{z-1}},  \notag \\
\lambda _{3} &=&\frac{1}{\sqrt{z-1}}.  \label{Constants}
\end{eqnarray}%
Using (\ref{Constants}), the metric function $f(r)$ may be written as%
\begin{eqnarray}
f(r) &=&\frac{kl^{2}}{r^{2}z^{2}}-\frac{2\Lambda _{4}l^{2}}{\left(
z+1\right) \left( z+2\right) }-\frac{m}{r^{z+2}}  \notag \\
&&+\frac{2^{p-1}q_{1}^{2\,p}{l}^{2\,p(z-1)+2}\left( 2p-1\right) {b}^{2(z-1)}%
}{\Gamma _{4}{r}^{\Gamma _{4}+z+2}}{,}
\end{eqnarray}%
where $\Gamma _{4}=z-2+2/(2p-1)$. Looking at $q_{3}^{2}$ in (\ref{Constants}%
), we find that the hyperbolic case $k=-1$ causes an imaginary charge except
for $z=1$. Therefore, we continue the paper just for the cases $k=0$ and $%
k=1 $. It is remarkable to note that our solutions include topological
asymptotic AdS black holes for $z=1$. At first glance the constants (\ref%
{Constants}) are infinite for $z=1$, but one should note that although $%
\lambda _{i}$s diverge at $z=1$, $\lambda _{i}\Phi $ is finite for this case
and $H_{i}=0$ and therefore action (\ref{action}) reduces to the Einstein
action in the presence of cosmological constant and nonlinear
electromagnetic field for $z=1$. This kind of solutions in the absence of
cosmological constant has been introduced in \cite{hasmart}. Also, this kind
of solutions in the presence of cosmological constant and in the context of
Lovelock gravity has been studied in \cite{MHM}.

In order to have asymptotic Lifshitz solutions, $f(r)$ should reduce to the
metric (\ref{Asym}). Thus, we have another restriction on $\Lambda _{4}$ as 
\begin{equation}
\Lambda _{4}=-\frac{(z+1)(z+2)}{2l^{2}}.  \label{Lambda}
\end{equation}%
So, the final form of the metric function is 
\begin{equation}
f(r)=1+\frac{kl^{2}}{r^{2}z^{2}}-\frac{m}{r^{z+2}}+\frac{q^{2p}}{{r}^{\Gamma
_{4}+z+2}}{,}  \label{Fr}
\end{equation}%
where $k=0,1$ and

\begin{equation}
q^{2p}=\frac{\left( 2p-1\right) {b}^{2(z-1)}}{2\Gamma _{4}{l}^{-2\,p(z-1)-2}}%
\left( 2q_{1}^{2}\right) ^{p}.  \label{q}
\end{equation}%
Using (\ref{FF}), (\ref{Phi}) and (\ref{Constants}), it is easy to show that
the gauge field with the free charge parameter is now given by%
\begin{equation*}
F_{rt}=\frac{q_{1}b^{2(z-1)}}{r^{\Gamma _{4}+1}},
\end{equation*}%
and therefore the gauge potential is 
\begin{equation}
A_{t}=-\frac{q_{1}b^{2(z-1)}}{\Gamma _{4}r^{\Gamma _{4}}}.  \label{At}
\end{equation}%
As we will show later in section (\ref{therm}), the dependence of total mass
of the black hole on $A_{t}$ is linear and therefore in order to have a
finite mass, $\Gamma _{4}$ should be positive. Thus, the metric function (%
\ref{Fr}) has the appropriate asymptotic given in Eq. (\ref{Asym}).

\subsection{PROPERTIES OF THE SOLUTIONS}

\begin{figure}[tbp]
\epsfxsize=7cm \centerline{\epsffile{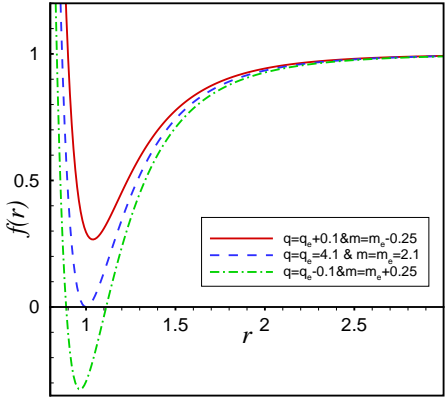}}
\caption{The function $f(r)$ versus $r$ for $p=0.8$, $z=3$, $l=b=1$, $k=0$
and $r_{\mathrm{ext}}=1$.}
\label{fig1}
\end{figure}

\begin{figure}[tbp]
\epsfxsize=7cm \centerline{\epsffile{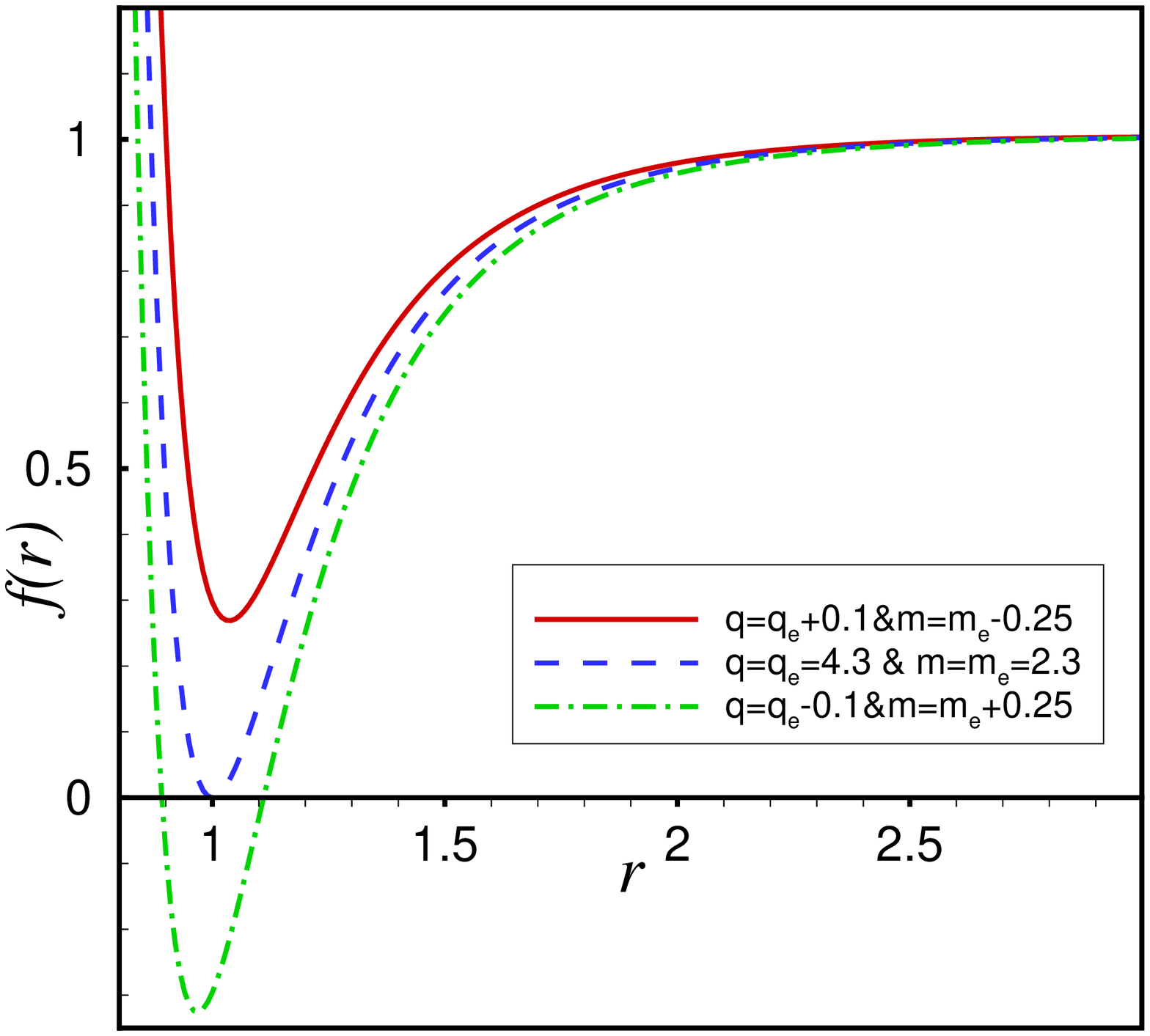}}
\caption{The function $f(r)$ versus $r$ for $p=0.8$, $z=3$, $l=b=1$, $k=1$
and $r_{\mathrm{ext}}=1$.}
\label{fig2}
\end{figure}

\begin{figure}[tbp]
\epsfxsize=7cm \centerline{\epsffile{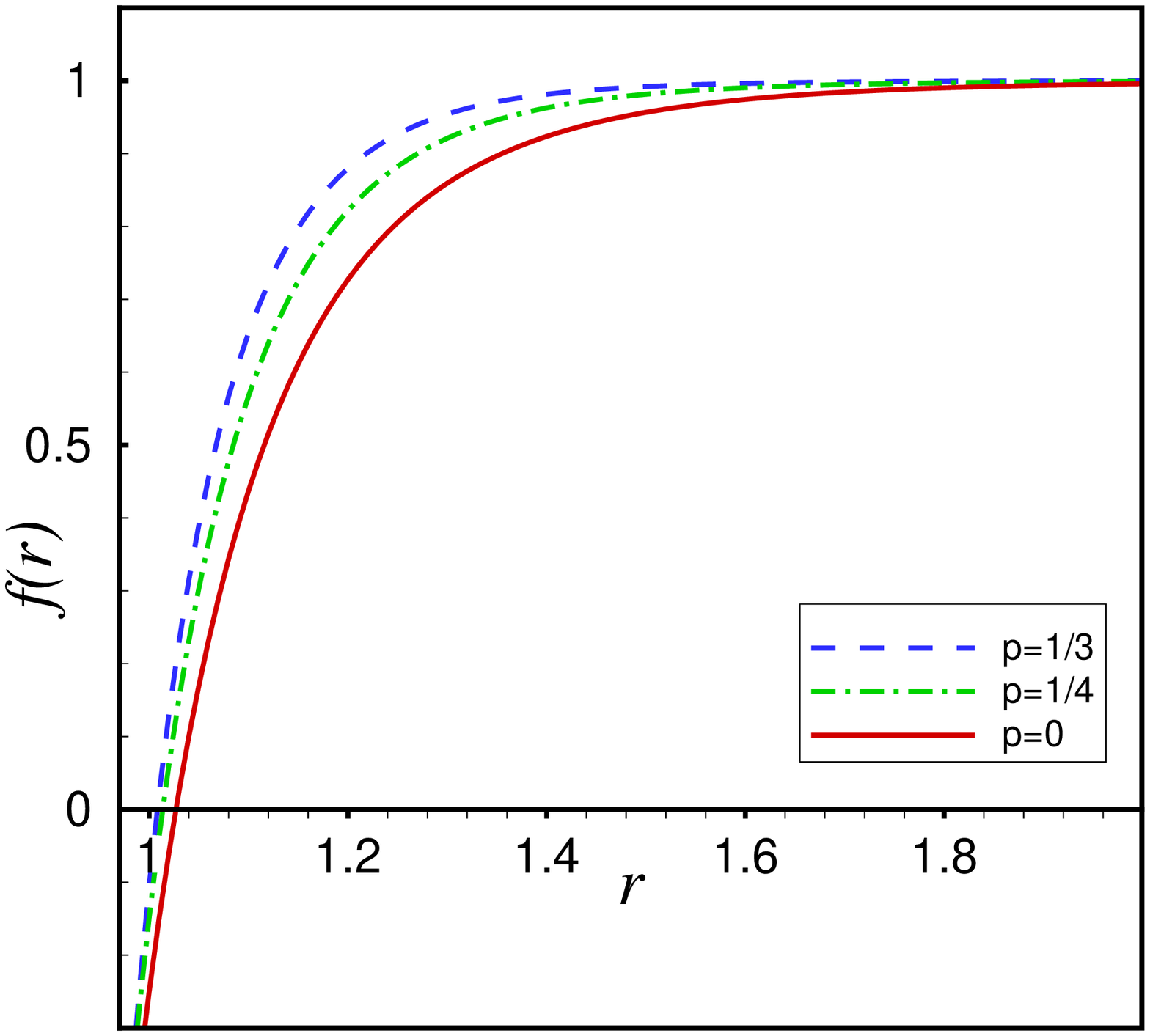}}
\caption{The function $f(r)$ versus $r$ for $p=1/3$ (dashed) $p=1/4$
(dashdot) $p=0$ (solid), $z=9$, $m=q=l=b=1$ and $k=0$.}
\label{fig3}
\end{figure}

\begin{figure}[tbp]
\epsfxsize=7cm \centerline{\epsffile{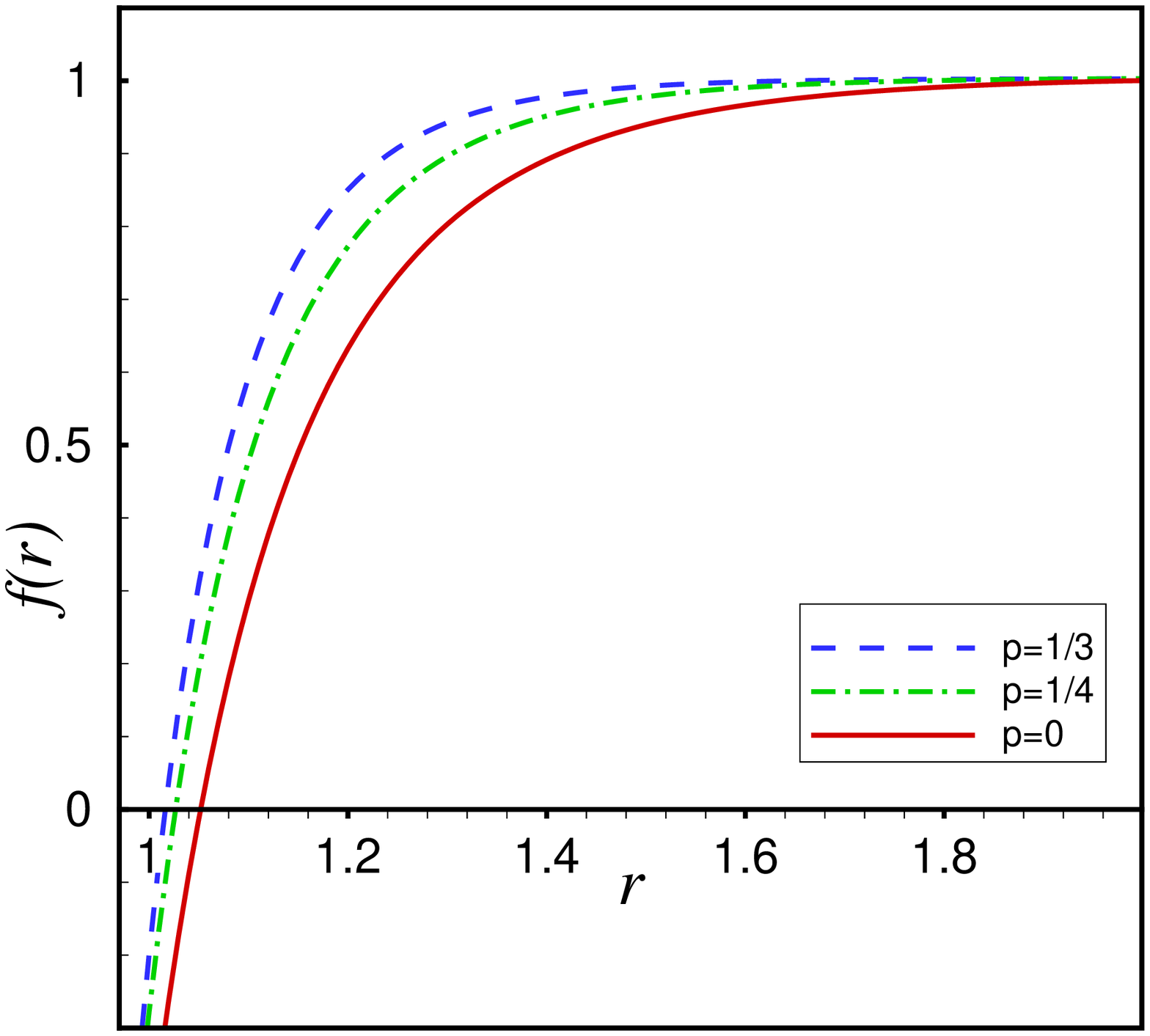}}
\caption{The function $f(r)$ versus $r$ for $p=1/3$ (dashed) $p=1/4$
(dashdot) $p=0$ (solid), $z=9$, $m=q=l=b=1$ and $k=1$.}
\label{fig4}
\end{figure}

In order to study the properties of the solutions, we first mention the
constraint on $z$ and $p$ which comes from the positivity of $\Gamma _{4}$: 
\begin{equation}
z-2+\frac{2}{(2p-1)}>0.  \label{Ups}
\end{equation}%
The inequality (\ref{Ups}) imposes the following restrictions on $p$ and $z$,

\begin{equation}
\begin{tabular}{ll}
$\text{for }p<1/2,$ & $z-1>(3-2p)/(1-2p),$ \\ 
&  \\ 
$\text{for }1/2<p\leq 3/2,$ & $\text{all }z(\geq 1)\text{ values are allowed,%
}$ \\ 
&  \\ 
$\text{for }p>3/2,$ & $z-1>(2p-3)/(2p-1).$%
\end{tabular}
\label{constraint}
\end{equation}%
For the allowed ranges of $p$ and $z$, the term including $q$ in $f(r)$ is
dominant as $r\rightarrow 0$. Therefore, as one can see from (\ref{Fr}) and (%
\ref{q}), since $\Gamma _{4}>0$ the sign of $f(r)$ in the vicinity of $r=0$
depends on the factor $2p-1$. Let us discuss the cases $p>1/2$ and $p<1/2$,
separately. For $p>1/2$, $f(r)$ is positive in the vicinity of $r=0$.
Therefore, one can have an extreme black hole with horizon radius $r_{+}=r_{%
\mathrm{ext}}$ provided $m=m_{\mathrm{ext}}$ or $q=q_{\mathrm{ext}}$, where%
\begin{equation}
q_{\mathrm{ext}}^{2p}=\frac{(z+2){r}_{\mathrm{ext}}^{\Gamma _{4}+z+2}}{%
\Gamma _{4}}+\frac{kl^{2}{r}_{\mathrm{ext}}^{\Gamma _{4}+z}}{\Gamma _{4}z}{,}
\end{equation}%
\begin{equation}
m_{\mathrm{ext}}=\frac{(\Gamma _{4}+z+2){r}_{\mathrm{ext}}^{z+2}}{\Gamma _{4}%
}+\frac{\left( \Gamma _{4}+z\right) kl^{2}{r}_{\mathrm{ext}}^{z}}{\Gamma
_{4}z^{2}}{.}
\end{equation}%
It is worthwhile to note that $m_{\mathrm{ext}}$ and $q_{\mathrm{ext}}$ are
calculated from equations $f^{\prime }(r_{\mathrm{ext}})=0$ and $f(r_{%
\mathrm{ext}})=0$, respectively. We have also solutions with two inner and
outer horizons $r_{-}<r_{\mathrm{ext}}<r_{+}$ provided $m>m_{\mathrm{ext}}$
or $q<q_{\mathrm{ext}}$ and naked singularities when $m<m_{\mathrm{ext}}$ or 
$q>q_{\mathrm{ext}}$ (See Figs. (\ref{fig1}) and (\ref{fig2})). For $p<1/2$, 
$f(r)\rightarrow -\infty $ as $r\rightarrow 0$ and therefore we have
Schwarzschild-like solutions. Note that in this case $\mathcal{L}_{\Phi }<0$%
\ and therefore this result matches with the general results we presented
before in this section. For this case, Figs. (\ref{fig3}) and (\ref{fig4})
depicts the behavior of $f(r)$ with suitable choices of $z$ (see Eq.(\ref%
{constraint})). 

\section{HIGHER-DIMENSIONAL LIFSHITZ BLACK HOLE/BRANE SOLUTIONS \label{ndim}}

In this section, we extend our solutions to the case of higher-dimensional
solutions. The action of ($n+1$)-dimensional theory ($n\geq 3$) can be
written as 
\begin{eqnarray}
S &=&-\frac{1}{16\pi }\int_{\mathcal{M}}d^{n+1}x\sqrt{-g}\left\{ \mathcal{R}-%
\frac{4}{n-1}(\nabla \Phi )^{2}\right.  \notag  \label{action2} \\
&&\left. -2\Lambda +\left( -e^{-4/(n-1)\lambda _{1}\Phi }F\right)
^{p}-\sum\limits_{i=2}^{3}e^{-4/(n-1)\lambda _{i}\Phi }H_{i}\right\} . 
\notag
\end{eqnarray}%
The variation of action (\ref{action2}) leads to the following field
equations 
\begin{eqnarray}
&&\mathcal{R}_{\mu \nu }=\frac{g_{\mu \nu }}{n-1}\left\{ 2\Lambda
+(2p-1)\left( -Fe^{-4\lambda _{1}\Phi /(n-1)}\right) ^{p}\right.  \notag \\
&&\left. -\sum\limits_{i=2}^{3}H_{i}e^{-4\lambda _{i}\Phi /(n-1)}\right\} +%
\frac{4}{n-1}\partial _{\mu }\Phi \partial _{\nu }\Phi  \notag \\
&&+2pe^{-4\lambda _{1}p\Phi /(n-1)}(-F)^{p-1}F_{\mu \lambda }F_{\nu }^{\text{
\ }\lambda }  \notag \\
&&+2\sum\limits_{i=2}^{3}e^{-4\lambda _{i}\Phi /(n-1)}\left( H_{i}\right)
_{\mu \lambda }\left( H_{i}\right) _{\nu }^{\text{ \ }\lambda },
\label{FFE1} \\
&&\nabla ^{2}\Phi -\frac{p{\lambda _{1}}}{2}e^{-{4\lambda _{1}p\Phi }/({n-1}%
)}(-F)^{p}  \notag \\
&&+\sum\limits_{i=2}^{3}\frac{{\lambda _{i}}}{2}e^{-{4\lambda _{i}\Phi }/({%
n-1})}H=0, \\
&&\triangledown _{\mu }\left( e^{-{4\lambda _{1}p\Phi }/({n-1}%
)}(-F)^{p-1}F^{\mu \nu }\right) =0, \\
&&\triangledown _{\mu }\left( e^{-{4\lambda _{i}\Phi }/({n-1})}\left(
H_{i}\right) ^{\mu \nu }\right) =0.  \label{FFE4}
\end{eqnarray}%
The line element of the higher-dimensional asymptotic Lifshitz spacetime is 
\begin{equation}
ds^{2}=-\frac{r^{2z}f(r)}{l^{2z}}dt^{2}+{\frac{l^{2}dr^{2}}{r^{2}f(r)}}%
+r^{2}d\Omega _{n-1}^{2},  \label{metric}
\end{equation}%
where $d\Omega _{n-1}^{2}$ is an ($n-1$)-dimensional hypersurface with
constant curvature $(n-1)(n-2)k$ and volume $\omega _{n-1}$. Following the
method of the previous section, one can find the solutions of the field
equations (\ref{FFE1})-(\ref{FFE4}) as 
\begin{eqnarray}
f(r) &=&1-{\frac{m}{{r}^{n-1+z}}}+{\frac{k{l}^{2}\left( n-2\right) ^{2}}{%
\left( z+n-3\right) ^{2}{r}^{2}}}+\frac{q^{2p}}{{r}^{\Gamma +z+n-1}},  \notag
\\
&&  \label{ff} \\
{\Phi (r)} &{=}&\frac{(n-1)\sqrt{z-1}}{2}{\ln \left( \frac{r}{b}\right) ,} \\
\left( A_{1}\right) _{t} &=&-\frac{q_{1}{b}^{2(z-1)}}{\Gamma {r}^{\Gamma }},
\\
\left( A_{2}\right) _{t} &=&\sqrt{\frac{z-1}{2(n+z-1)}}{\frac{{r}^{n+z-1}}{{l%
}^{z}{b}^{n-1}}}, \\
\left( A_{3}\right) _{t} &=&{\frac{\sqrt{{k\left( n-1\right) \left(
n-2\right) \left( z-1\right) }}{r}^{z+n-3}\,}{\sqrt{2}(z+n-3)^{3/2}{l}^{z-1}{%
b}^{n-2}}},
\end{eqnarray}%
where 
\begin{eqnarray*}
\Gamma &=&z-2+(n-1)/(2p-1), \\
q^{2p} &=&\frac{\left( 2\,p-1\right) {b}^{2(z-1)}}{\left( n-1\right) {l}%
^{-2\,p\left( z-1\right) -2}\Gamma }\left( 2q_{1}^{2}\right) ^{p},
\end{eqnarray*}%
and $\Lambda =-(z+n-1)(z+n-2)/2l^{2}$. Here we pause to point out some
remarks as in the previous sections. First, $(A_{3})_{t}$ is imaginary in
the case of $k=-1$ and therefore our solutions rule out the case with
hyperbolic horizon. Second, $A_{t}$ should be finite at infinity since the
total mass linearly depends on it as we will see later in section (\ref%
{therm}). Therefore $\Gamma $ should be positive. This constraint makes some
limits on $z$ and $p$. The allowed ranges for $p$ and $z$ are:%
\begin{equation}
\begin{tabular}{ll}
$\text{for }p<1/2,$ & $z-1>(n-2p)/(1-2p),$ \\ 
&  \\ 
$\text{for }1/2<p\leq n/2,$ & $\text{all }z(\geq 1)\text{ values are allowed,%
}$ \\ 
&  \\ 
$\text{for }p>n/2,$ & $z-1>(2p-n)/(2p-1).$%
\end{tabular}
\label{constraint2}
\end{equation}%
For these allowed ranges of $p$ and $z$, the term proportional to charge in
the metric function disappears at infinity and it dominates as $r\rightarrow
0$. Therefore, since this term is positive for $p>1/2$, in this case one may
encounter a black hole with inner and outer horizons or extreme black hole
for suitable choices of $m$ and $q$ where

\begin{eqnarray}
q_{\mathrm{ext}}^{2p} &=&\frac{{(z+n-1)}}{\Gamma }{{r}_{\mathrm{ext}%
}^{z+n-1+\Gamma }}  \notag \\
&&{+}\frac{\left( n-2\right) ^{2}k{l}^{2}}{\Gamma \left( z+n-3\right) }{r}_{%
\mathrm{ext}}^{z+n-3+\Gamma }, \\
{m}_{\mathrm{ext}} &=&\frac{{(\Gamma +z+n-1)}}{\Gamma }{{r}_{\mathrm{ext}%
}^{z+n-1}}  \notag \\
&&{+}\frac{\left( \Gamma +z+n-3\right) \left( n-2\right) ^{2}k{l}^{2}}{%
\Gamma \left( z+n-3\right) ^{2}}{r}_{\mathrm{ext}}^{z+n-3}.
\end{eqnarray}%
On the other hand, for $p<1/2$ where charge term is negative in metric
function, we have Schwarzschild-like solutions.

\section{THERMODYNAMICS OF LIFSHITZ BLACK HOLES/BRANES \label{therm}}

In this section, we seek for satisfaction of the first law of thermodynamics
for our Lifshitz black hole/brane solutions. We start with the calculation
of mass that is fundamental for thermodynamics discussions. We use the
modified subtraction method of Brown and York (BY) \cite{BY}. In order to
use the modified BY method \cite{modBY}, the metric should be written in the
form%
\begin{equation}
ds^{2}=-X({\mathcal{R}})dt^{2}+\frac{d\mathcal{R}^{2}}{Y(\mathcal{R})}+%
\mathcal{R}^{2}d\Omega ^{2}.  \label{Mets}
\end{equation}%
It is obvious from (\ref{metric}) that in our case $\mathcal{R}=r$ and
therefore%
\begin{equation}
X(\mathcal{R})=Y(\mathcal{R)}=f(r(\mathcal{R})).
\end{equation}%
The background metric is chosen to be the metric (\ref{Mets}) with

\begin{equation}
X_{0}(\mathcal{R})=Y_{0}(\mathcal{R})=f_{0}(r(\mathcal{R}))=1.
\end{equation}%
The quasilocal conserved mass can be calculated via

\begin{eqnarray}
M &=&\frac{1}{8\pi }\int_{\mathcal{B}}d^{2}\varphi \sqrt{\sigma }\left\{
\left( K_{ab}-Kh_{ab}\right) \right.  \notag \\
&&\left. -\left( K_{ab}^{0}-K^{0}h_{ab}^{0}\right) \right\} n^{a}\xi ^{b},
\end{eqnarray}%
where $\sigma $ is the determinant of the metric of the boundary $\mathcal{B}
$, $K_{ab}^{0}$ is the extrinsic curvature of the background metric and $%
n^{a}$ and $\xi ^{b}$ are the timelike unit normal vector to the boundary $%
\mathcal{B}$ and a timelike Killing vector field on the boundary surface,
respectively. Using the above modified BY formalism, one can calculate the
mass of the space time per unit volume ${\omega _{n-1}}$ as%
\begin{equation}
M=\frac{(n-1)m}{16\pi l^{z+1}},  \label{Mass}
\end{equation}%
where the mass parameter $m$ can be written in term of the horizon radius $%
r_{+}$ by using the fact that $f(r_{+})=0$: 
\begin{equation}
{m(r_{+})}={r}_{+}^{z+n-1}+{\frac{k{l}^{2}\left( n-2\right) ^{2}{r}%
_{+}^{z+n-3}}{\left( z+n-3\right) ^{2}}}+\frac{q^{2p}}{{r}_{+}^{\Gamma }}.
\label{mr+}
\end{equation}%
One can also calculate the charge by using the Gauss law 
\begin{equation}
Q=\frac{\,{1}}{4\pi }\int r^{n-1}e^{-{4\lambda _{1}p\Phi /(n-1)}%
}(-F)^{p-1}F_{\mu \nu }n^{\mu }u^{\nu }d{\Omega },  \label{chdef}
\end{equation}%
where $n^{\mu }$ and $u^{\nu }$ are the unit spacelike and timelike normals
to the hypersurface of radius $r$ given as 
\begin{gather*}
n^{\mu }=\frac{1}{\sqrt{-g_{tt}}}dt=\frac{l^{z}}{r^{z}\sqrt{f(r)}}dt, \\
u^{\nu }=\frac{1}{\sqrt{g_{rr}}}dr=\frac{r\sqrt{f(r)}}{l}dr.
\end{gather*}%
Using (\ref{chdef}), we obtain the charge per unit volume $\omega _{n-1}$ as 
\begin{equation}
Q=\frac{2^{p-1}\left( q_{1}l^{z-1}\right) ^{2p-1}}{4\pi }.  \label{charge}
\end{equation}%
The electric potential $U$, measured at infinity with respect to horizon is
defined by 
\begin{equation}
U=A_{\mu }\chi ^{\mu }\left\vert _{r\rightarrow \infty }-A_{\mu }\chi ^{\mu
}\right\vert _{r=r_{+}},  \label{Pot}
\end{equation}%
where $\chi =C\partial _{t}$ is the null generator of the horizon. The value
of constant $C$ will be explained later. Using (\ref{At}), we can obtain
electric potential 
\begin{equation}
U=\frac{Cq_{1}b^{2(z-1)}}{\Gamma r_{+}^{\Gamma }}.  \label{elecpot}
\end{equation}%
The entropy of the black holes can be calculated by using the area law of
the entropy which is applied to almost all kinds of black holes in Einstein
gravity including dilaton black holes \cite{Beck,hunt}. Thus, the entropy of
our solutions per unit volume $\omega _{n-1}$ is 
\begin{equation}
S=\frac{r_{+}^{n-1}}{4}.  \label{entropy}
\end{equation}%
The Hawking temperature can also be obtained as 
\begin{eqnarray}
T_{+} &=&\frac{r_{+}^{z+1}f^{\prime }\left( r_{+}\right) }{4\pi l^{z+1}} 
\notag \\
&=&\frac{1}{4\pi }\left\{ \frac{{(n-1+z){r}_{+}^{z}}}{l^{z+1}}+{\frac{%
k\left( n-2\right) ^{2}{r}_{+}^{z-2}}{l^{z-1}\left( z+n-3\right) }}-\frac{%
\Gamma q^{2p}}{l^{z+1}{r}_{+}^{\Gamma +n-1}}\right\} .  \notag \\
&&  \label{Temp}
\end{eqnarray}%
In order to check the first law of thermodynamics, we should write a
Smarr-type formula. Using (\ref{Mass}), (\ref{mr+}), (\ref{charge}) and (\ref%
{entropy}), we can write the Smarr-type formula as 
\begin{eqnarray}
M\left( S,Q\right) &=&\frac{(n-1)\left( 4S\right) ^{(n-1+z)/(n-1)}}{16\pi
l^{z+1}}  \notag \\
&&+{\frac{k(n-1)\left( n-2\right) ^{2}\left( 4S\right) ^{(z+n-3)/(n-1)}}{%
16\pi l^{z-1}\left( z+n-3\right) ^{2}}}  \notag \\
&&+\frac{\left( 2\,p-1\right) \left( \pi Q\right) ^{2p/(2p-1)}\left(
4S\right) ^{-\Gamma /(n-1)}}{\pi {l}^{z-1}2^{\left( 3p-4\right)
/(2p-1)}\Gamma {b}^{2(1-\,z)}}.  \notag \\
&&
\end{eqnarray}%
We can now consider $S$ and $Q$ as a complete set of extensive quantities
for mass $M(S,Q)$ and define temperature $T$ and electric potential $U$ as
their conjugate intensive quantities, respectively: 
\begin{equation}
T=\left( \frac{\partial M}{\partial S}\right) _{Q}\text{ \ \ \ \ and \ \ \ \ 
}U=\left( \frac{\partial M}{\partial Q}\right) _{S}.  \label{intqua}
\end{equation}%
Calculations show that intensive quantities calculated by (\ref{intqua})
coincide with those computed by (\ref{elecpot}) and (\ref{Temp}) provided $%
C=p$. Therefore, these thermodynamics quantities satisfy the first law of
thermodynamics:%
\begin{equation}
dM=TdS+UdQ.  \label{TFL}
\end{equation}%
According to (\ref{TFL}), it is clear that total mass linearly depends on $U$%
. On the other hand $U$ is proportional to $A_{t}$ (see Eqs. (\ref{Pot}) and
(\ref{elecpot})). Therefore $A_{t}$ should be finite at infinity in order to
have a finite mass as we have applied this condition on our solutions (see
Eq. (\ref{constraint2})).

\section{THERMAL STABILITY IN THE CANONICAL AND GRAND-CANONICAL ENSEMBLES}

\label{stab} 
\begin{figure}[tbp]
\epsfxsize=7cm \centerline{\epsffile{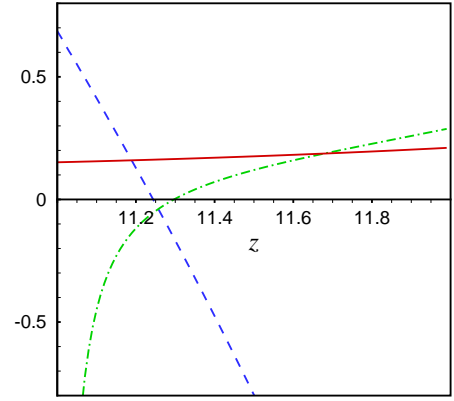}}
\caption{The behavior of $10T$ (solid), $10(\partial ^{2}M/\partial
S^{2})_{Q}$ (dashed) and $\mathbf{H}_{S,Q}^{M}$ (dashdot) versus $z$ for $%
k=0 $ with $l=b=1$, $q_{1}=0.1$, $r_{+}=0.6$, $n=4$ and $p=1/3$.}
\label{fig5}
\end{figure}
\begin{figure}[tbp]
\epsfxsize=7cm \centerline{\epsffile{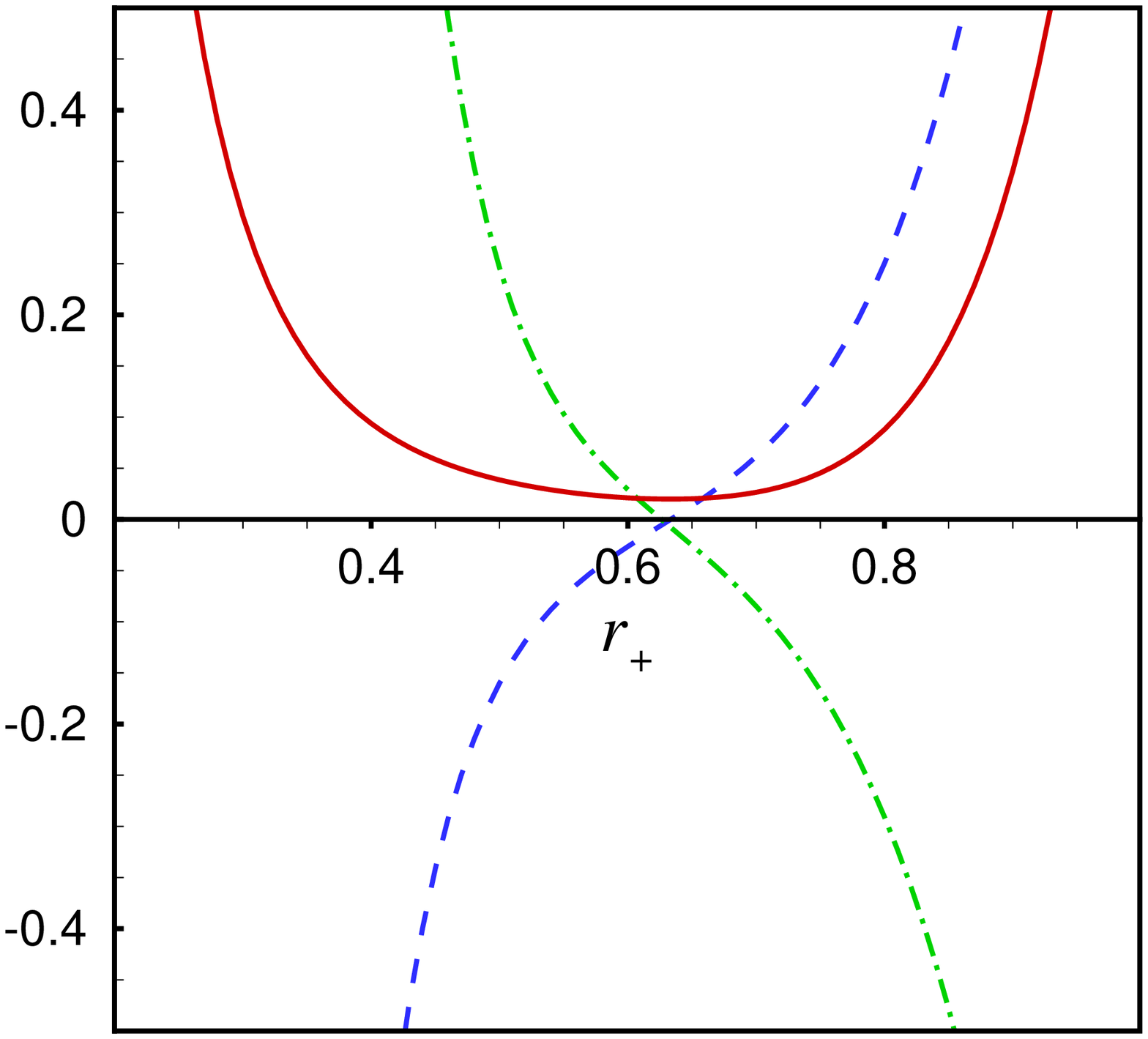}}
\caption{The behavior of $T$ (solid), $10^{-1}(\partial ^{2}M/\partial
S^{2})_{Q}$ (dashed) and $10^{-1}\mathbf{H}_{S,Q}^{M}$ (dashdot) versus $%
r_{+}$ for $k=0$ with $l=b=1$, $q_{1}=0.1$, $z=12$, $n=4$ and $p=1/3$.}
\label{fig6}
\end{figure}
\begin{figure}[tbp]
\epsfxsize=7cm \centerline{\epsffile{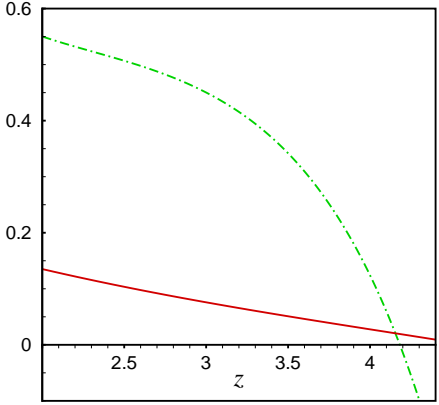}}
\caption{The behavior of $T$ (solid) and $10^{-3}\mathbf{H}_{S,Q}^{M}$
(dashdot) versus $z$ for $k=0$ with $l=b=1$, $q_{1}=0.2$, $r_{+}=0.55$, $n=5$
and $p=2$.}
\label{fig7}
\end{figure}
\begin{figure}[tbp]
\epsfxsize=7cm \centerline{\epsffile{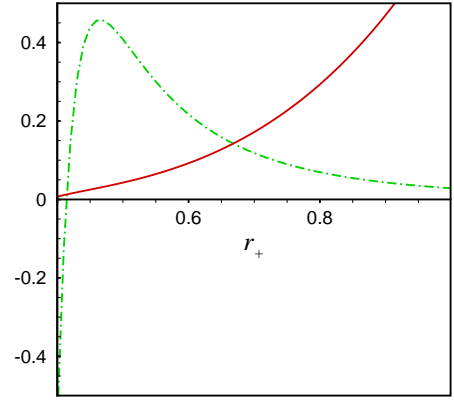}}
\caption{The behavior of $T$ (solid) and $10^{-5}\mathbf{H}_{S,Q}^{M}$
(dashdot) versus $r_{+}$ for $k=0$ with $l=b=1$, $q_{1}=0.15$, $z=4$, $n=6$
and $p=3$.}
\label{fig8}
\end{figure}
\begin{figure}[tbp]
\epsfxsize=7cm \centerline{\epsffile{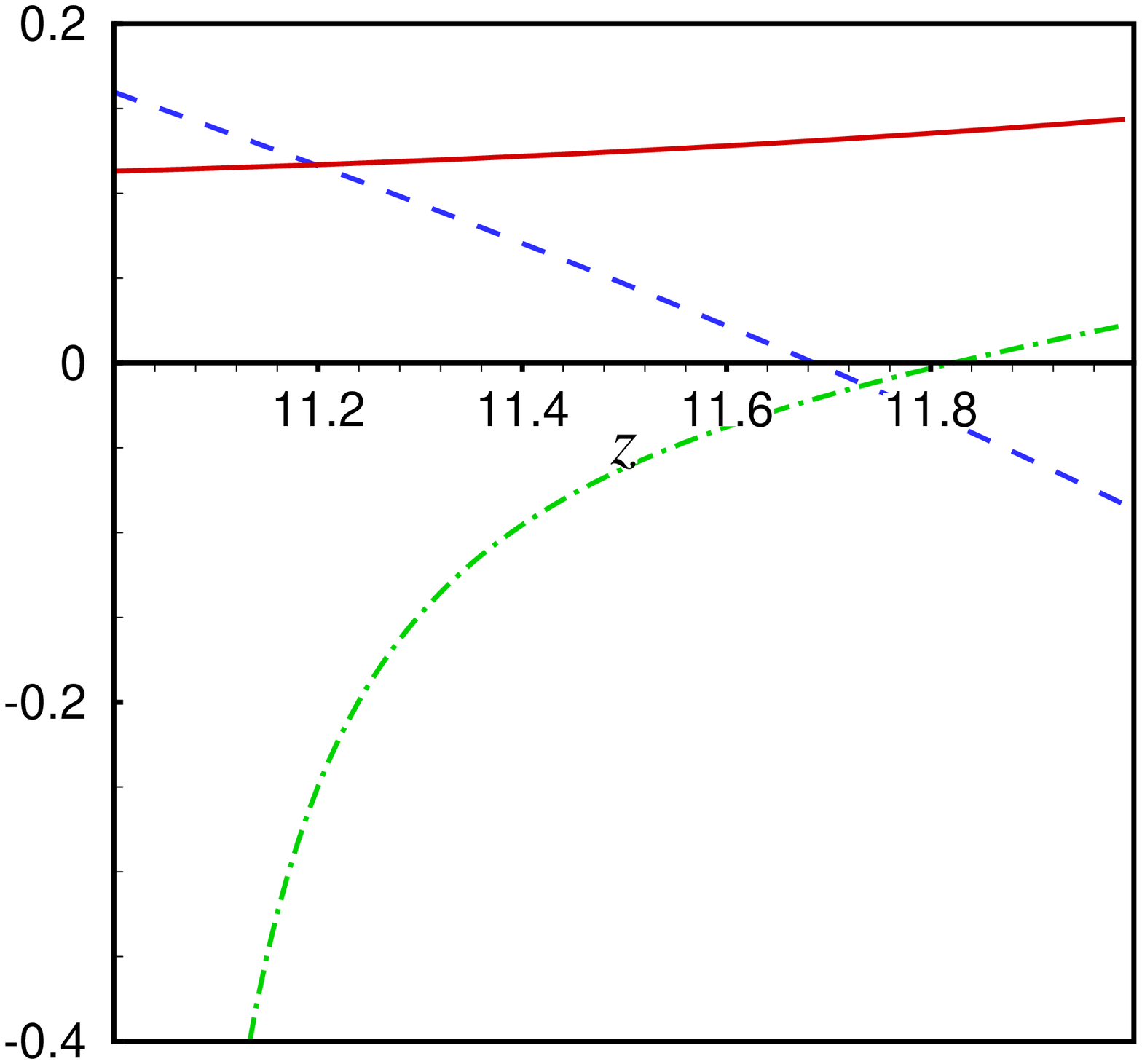}}
\caption{The behavior of $10T$ (solid), $(\partial ^{2}M/\partial S^{2})_{Q}$
(dashed) and $\mathbf{H}_{S,Q}^{M}$ (dashdot) versus $z$ for $k=1$ with $%
l=b=1$, $q_{1}=0.05$, $r_{+}=0.6$, $n=4$ and $p=1/3$.}
\label{fig9}
\end{figure}
\begin{figure}[tbp]
\epsfxsize=7cm \centerline{\epsffile{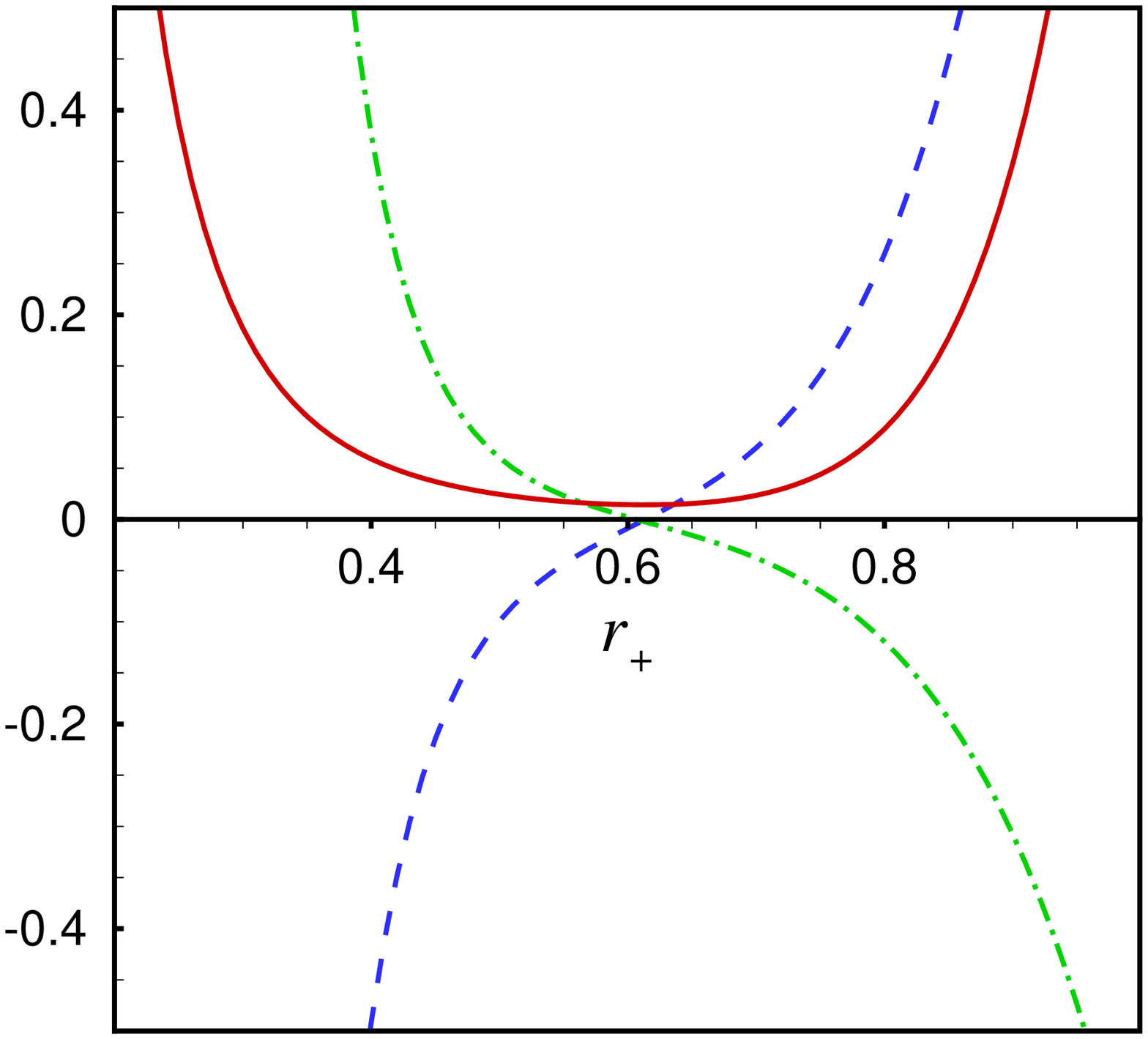}}
\caption{The behavior of $T$ (solid), $10^{-1}(\partial ^{2}M/\partial
S^{2})_{Q}$ (dashed) and $10^{-1}\mathbf{H}_{S,Q}^{M}$ (dashdot) versus $%
r_{+}$ for $k=1$ with $l=b=1$, $q_{1}=0.05$, $z=12$, $n=4$ and $p=1/3$.}
\label{fig10}
\end{figure}
\begin{figure}[tbp]
\epsfxsize=7cm \centerline{\epsffile{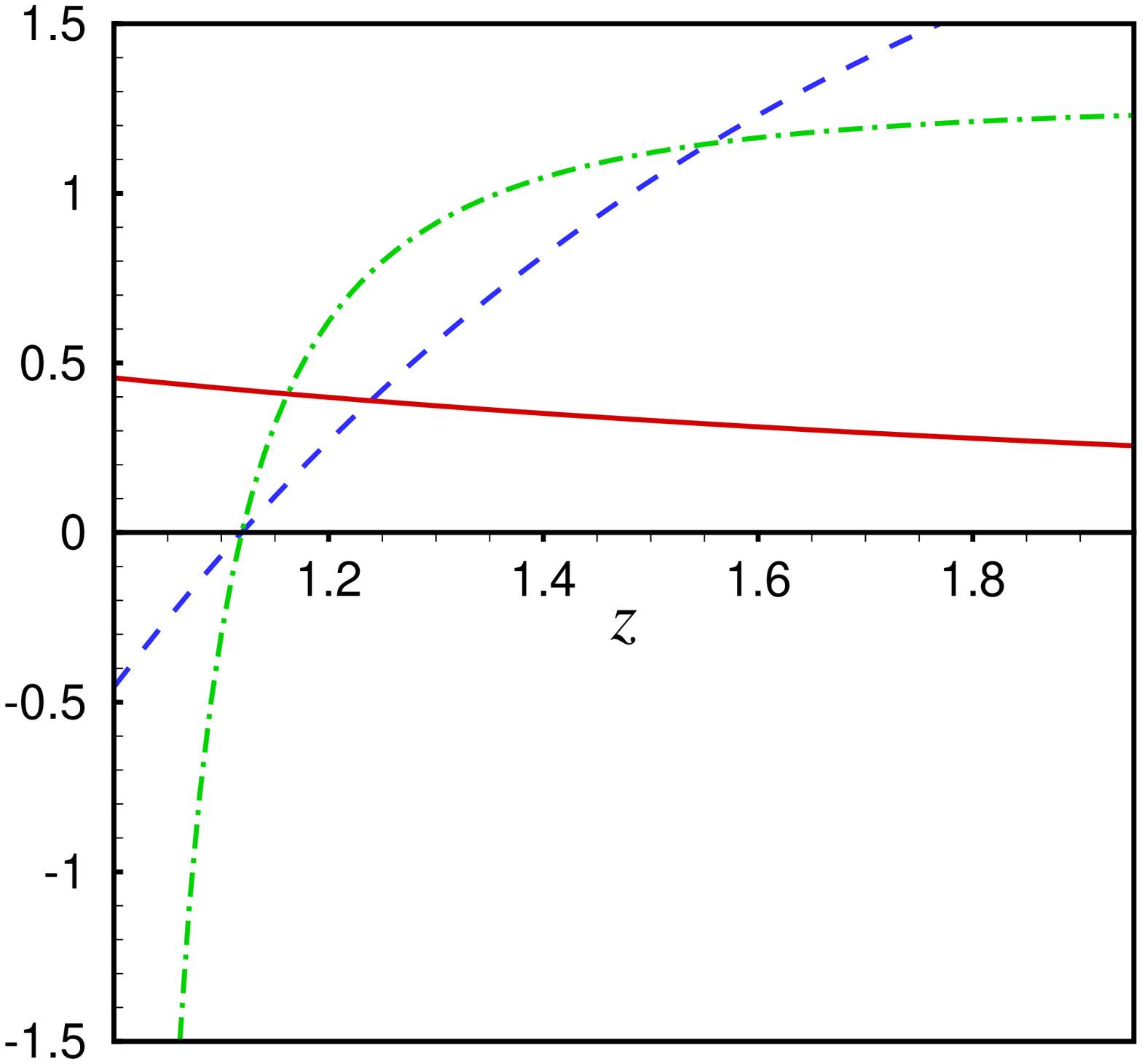}}
\caption{The behavior of $T$ (solid), $(\partial ^{2}M/\partial S^{2})_{Q}$
(dashed) and $10^{-3}\mathbf{H}_{S,Q}^{M}$ (dashdot) versus $z$ for $k=1$
with $l=b=1$, $q_{1}=0.1$, $r_{+}=0.6$, $n=4$ and $p=2$.}
\label{fig11}
\end{figure}
\begin{figure}[tbp]
\epsfxsize=7cm \centerline{\epsffile{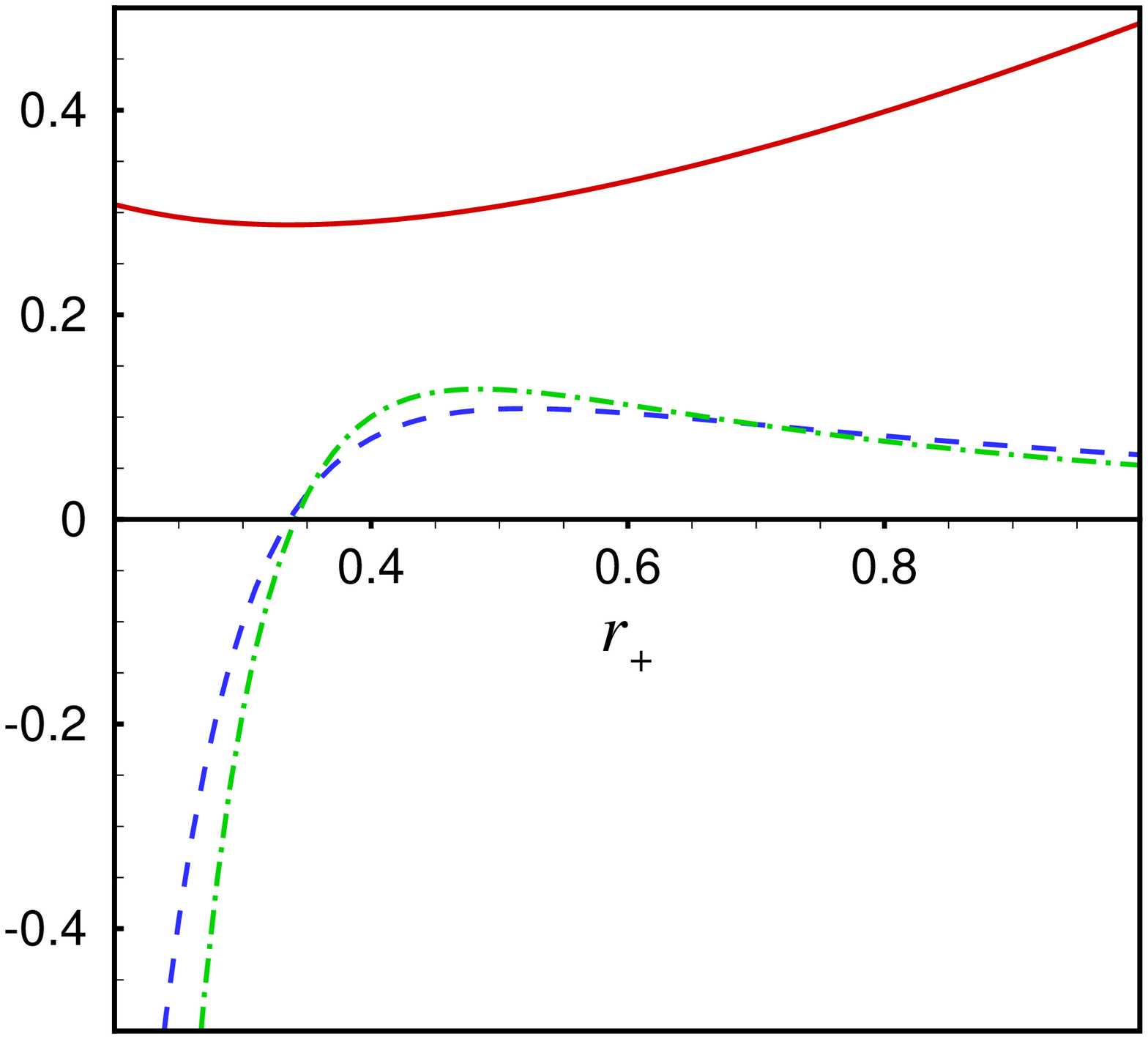}}
\caption{The behavior of $T$ (solid), $10^{-1}(\partial ^{2}M/\partial
S^{2})_{Q}$ (dashed) and $10^{-4}\mathbf{H}_{S,Q}^{M}$ (dashdot) versus $%
r_{+}$ for $k=1$ with $l=b=1$, $q_{1}=0.1$, $z=1.5$, $n=4$ and $p=2$.}
\label{fig12}
\end{figure}
In order to study thermal stability of the asymptotic Lifshitz dilaton black
solutions, we adopt two different ensembles, namely, canonical and
grand-canonical ensembles. In the canonical ensemble, the charge is a fixed
parameter and therefore the positivity of the heat capacity $%
C_{v}=T/(\partial ^{2}M/\partial S^{2})_{Q}$ is sufficient to ensure the
local stability \cite{Cal2,Gub}. This implies that the positivity of $%
(\partial ^{2}M/\partial S^{2})_{Q}$ guarantees the local stability of the
solutions in the ranges where $T$ is positive. It is a matter of
calculations to show that 
\begin{eqnarray}
\left( \frac{\partial ^{2}M}{\partial S^{2}}\right) _{Q} &=&\frac{%
z(n-1+z)r_{+}^{z-n+1}}{(n-1)\pi l^{z+1}}  \notag \\
&&+{\frac{k\left( n-2\right) ^{2}(z-2)r_{+}^{z-n-1}}{\pi l^{z-1}(n-1)\left(
z+n-3\right) }}  \notag \\
&&+\frac{2^{p}(\Gamma +n-1)\left( 2\,p-1\right) q_{1}^{2p}r_{+}^{-\Gamma
-2n+2}}{\pi {l}^{(1-2p)(z-1)}(n-1)^{2}{b}^{2(1-\,z)}}.  \notag \\
&&  \label{dMSS}
\end{eqnarray}%
In grand-canonical ensemble $Q$ is no longer fixed. In our case, the mass is
a function of entropy and charge and therefore the system is locally stable
provided $\mathbf{H}_{S,Q}^{M}=\left[ \partial ^{2}M/\partial S\partial Q%
\right] >0$ where the determinant of Hessian matrix can be calculated as 
\begin{eqnarray}
\mathbf{H}_{S,Q}^{M} &=&\frac{pz(n-1+z)l^{2p(1-z)-2}r_{+}^{-\Gamma +z-n+1}}{%
2^{p-3}(2p-1)\Gamma {b}^{2(1-\,z)}(n-1)q_{1}^{2(p-1)}}  \notag \\
&&+{\frac{pk\left( n-2\right) ^{2}(z-2){b}^{2(z-1)}l^{2p\left( 1-z\right)
}r_{+}^{-\Gamma +z-n-1}}{2^{p-3}(2p-1)(n-1)\left( z+n-3\right) \Gamma
q_{1}^{2(p-1)}}}  \notag \\
&&-\frac{8p\left( 2p-1\right) \left( z-2\right) q_{1}^{2}r_{+}^{-2\Gamma
-2n+2}}{(n-1)^{2}\Gamma {b}^{4(1-\,z)}}.  \label{Hess}
\end{eqnarray}%
Let us now discuss thermal stability of the solutions in both ensembles. As
it is obvious from (\ref{dMSS}) and (\ref{Hess}) for $z=2$ where the allowed
range of $p$ is $p>1/2$ (see Eq.(\ref{constraint2})), we have thermally
stable solutions in both canonical and grand-canonical ensembles. For other
ranges of the parameters, we consider the case $k=0$ and $k=1$ separately:

$i$) $k=0$: As one can see from Fig. \ref{fig5}, for $p<1/2$ the behaviors
of $\left( \partial ^{2}M/\partial S^{2}\right) _{Q}$ and $\mathbf{H}%
_{S,Q}^{M}$ in terms of $z$ are opposite. That is in canonical ensemble,
there is a $z_{\max }$ such that the solutions are stable provided $%
z<z_{\max }$, whereas in the grand-canonical ensemble there is a $z_{\min }$
in which for values greater than it we have thermally stable solutions.
Also, in this range there is a $r_{+\min }$ in the canonical ensemble that
for values greater than it we have stable solutions while in the
grand-canonical ensemble there is a $r_{+\max }$ that for radius of horizon
greater than it one encounters unstable solutions. This behavior is shown in
Fig. \ref{fig6}. Of course, one should note that for $p<0$ where $z$ is
always greater than $2$ (Eq. (\ref{constraint2})), there are stable
solutions in grand-canonical ensemble. We also encounter stable solutions
for $p>1/2$ for all $z(\geq 1)$ values in canonical ensemble while the
solutions are stable in the grand-canonical ensemble just for $1\leq z<2$.
In the grand-canonical ensemble when $p>1/2$ and $z>2$ there are unstable
solutions for $z>z_{\max }>2$ (Fig. \ref{fig7}). In latter range there is $%
r_{+\min }$ that for values greater than it we have stable black branes
(Fig. \ref{fig8}).

$ii$) $k=1$: In this case, solutions are stable in grand-canonical ensemble
for $p<0$ whereas for $0<p<1/2$ there are unstable solutions provided $%
z<z_{\min }$ (Fig. \ref{fig9}). As one can see in Fig. \ref{fig9}, in
canonical ensemble, there are stable black holes provided $z<z_{\max }$ when 
$p<1/2$. In latter range there is an $r_{+\max }$ in grand-canonical
ensemble that for $r_{+}<r_{+\max }$ black holes are stable while in
canonical ensemble black holes with horizon radius lower than $r_{+\min }$
are unstable (Fig. \ref{fig10}). Black holes are always stable in canonical
ensemble if $p>1/2$ and $z>2$. For $p>1/2$, we also have stable solutions
for $1\leq z<2$ in canonical ensemble provided $z>z_{\min }$ as it is clear
from Fig. \ref{fig11}. In grand-canonical ensemble there is also a $z_{\min
} $ that for values lower than it solutions are unstable for $p>1/2$ (Fig. %
\ref{fig11}). In later range black holes in both canonical and
grand-canonical ensembles are stable provided that their radius of horizon $%
r_{+}>r_{+\min }$ (See Fig. \ref{fig12}).

\section{DYNAMICAL STABILITY OF 4-DIMENSIONAL AdS BLACK HOLES \label{dystab}}

In previous section we investigated the termal stability of topological
black hole solutions. In addition to thermal stability, it is worth
discussing stability of black hole solutions under dynamical perturbations.
Therefore, in this section, we intend to discuss dynamical stability of
4-dimensional AdS black hole solutions. It has been shown that the
properties of perturbations of a 4-dimensional static and spherically
symmetric background under a two-dimensional rotation transformation can be
decomposed into odd- and even-parity sectors \cite{regwhe}. These
perturbations can also be written in terms of the sum of spherical harmonics 
$Y_{m}^{\ell }$\ and the dynamical stability is examined by studying the
behavior of perturbation modes.

Dynamical stability of black hole solutions in Einstein gravity coupled to
nonlinear electrodynamics (NED) sources is presented in \cite{nedsta}. The
Hamiltonian of a NED Lagrangian $\mathcal{L}(\hat{F})$\ where $\hat{F}=F/4$,
can be written as $\mathcal{H(}\hat{F})\equiv 2\mathcal{L}_{\hat{F}}\hat{F}-%
\mathcal{L}$. It is convenient to study the dynamical stability in so-called 
$P$\ frame where $P=\mathcal{L}_{\hat{F}}^{2}\hat{F}$\ and $\mathcal{H(}\hat{%
F})\rightarrow \mathcal{H}(P)$. In this frame, there are dynamically stable
solutions under odd-type perturbations provided $\mathcal{H}_{P}$\ has no
zero outside the horizon. For even-type perturbations, we have dynamically
unstable black hole solutions provided $\mathcal{H}_{xx}>0$\ where $x=\sqrt{%
-2Q^{2}\,P}$. In the case of AdS solutions ($z=1$), one can calculate

\begin{equation}
\mathcal{H}_{P}=\frac{1}{p}\left( \frac{-4P}{p^{2}}\right) ^{\frac{1-p}{2p-1}%
},  \label{Hp}
\end{equation}%
where $P=-p^{2}\left( 2F_{tr}^{2}\right) ^{2p-1}/4$. Since $\mathcal{H}_{P}$%
\ has no root outside the horizon, we have stable solutions under odd-type
perturbations. $\mathcal{H}_{xx}$\ can also be calculated as

\begin{equation}
\mathcal{H}_{xx}=-\frac{1}{p\left( 2p-1\right) Q^{2}}\left( \frac{2x^{2}}{%
p^{2}Q^{2}}\right) ^{\frac{1-p}{2p-1}}.  \label{Hxx}
\end{equation}%
Therefore, we encounter unstable solutions under even-type perturbations
provided $0<p<1/2$.

\section{CLOSING REMARKS}

In this paper, we constructed a new class of asymptotically Lifshitz
topological black hole solutions of Einstein-dilaton gravity coupled to a
power-law Maxwell field. The motivation for taking this kind of Lagrangian
originates from the fact that Maxwell Lagrangian is conformally invariant
only in four dimensions. In order to build a version of Maxwell theory which
enjoy the conformally invariance property in all higher dimensions, the
so-called power-law Maxwell electrodynamics was proposed in Ref. \cite%
{hassaine}. In fact, in ($n+1$)-dimensional spacetime the Lagrangian $\left[
-e^{-4/(n-1)\lambda \Phi }F_{\mu \nu }F^{\mu \nu }\right] ^{p}$ is invariant
under the conformal transformation $g_{\mu \nu }\rightarrow \Omega
^{2}g_{\mu \nu }$, $A_{\mu }\rightarrow A_{\mu }$ and $\Phi \rightarrow \Phi 
$, provided $p=(n+1)/4$. We first considered the four dimensional action and
by using this toy model we explained the procedure of obtaining the
asymptotic Lifshitz topological charged black holes of the theory in detail.
We allowed the horizon to be a hypersurface with $k=1,0,-1$ constant
curvature. Although we found that the reality of the charge parameter rules
out the case $k=-1$\textbf{\ }in the presence of dilaton and two
electromagnetic fields, a topological Lifshitz black hole with $k=-1$\textbf{%
\ }has been obtained in Ref. \cite{Mann} in the presence of a massive
electromagnetic field. Indeed, in the solution obtained in Ref. \cite{Mann},
the massive electromagnetic field makes the asymptotic behavior to be
Lifshitz, while here the set of dilaton and two linear electromagnetic
fields guarantees the asymptotic Lifshitz behavior for topological black
hole. We also discussed the possible ranges of the parameters for which the
metric function goes to $f(r)\rightarrow 1+kl^{2}/(z^{2}r^{2})$ as $%
r\rightarrow \infty $ where $z$ is a dynamical critical exponent shows the
degrees of anisotropy between space and time in Lifshitz spacetime. Then, we
generalized our study to all higher dimensions. We found that our ($n+1$%
)-dimensional solutions are asymptotically Lifshitz provided for $p<1/2$ and 
$p>n/2$, $z-1>(2p-n)/(2p-1)$ while for $1/2<p\leq n/2$, all $z(\geq 1)$
values are allowed. Using the modified Brown York method, we computed the
mass of the solutions. Next, by computing the thermodynamical quantities
associated with $(n+1)$-dimensional asymptotic Lifshitz dilaton black
holes/branes with power-law Maxwell field and obtaining the Smarr formula $%
M(S,Q)$, we checked the validity of the first law of thermodynamics, $%
dM=TdS+UdQ$, on the horizon. We also explore thermal stability of the
obtained solutions both in canonical and grand-canonical ensembles. We found
that our solutions are always stable in both canonical and grand-canonical
ensembles for $z=2$ where the allowed range of $p$ is $p>1/2$. In
grand-canonical ensemble, our black hole/brane solutions are always stable
for $p<0.$ In this ensemble, there are stable solutions for $0<p<1/2$
provided $z>z_{\min }$ or $r<r_{+\max }$. For $p>1/2$, black branes are
stable for $1\leq z<2$ while they are stable for $z<z_{\max }$ or $%
r>r_{+\min }$ when $z>2$. In latter range of $p$, black holes are stable
provided $z(r_{+})>z_{\min }(r_{+\min })$. In canonical ensemble, our black
hole/brane solutions encounter instability for $p<1/2$ and $z>z_{\max }$ or $%
r<r_{+\min }$. For $p>1/2$, black branes ($k=0$) are always stable in
canonical ensemble while black holes ($k=1$) are stable for $z>2$. In latter
range of $p$, black hole solutions encounter instability in canonical
ensemble in the range of $1\leq z<2$ provided $z<z_{\min }$ or $%
r_{+}<r_{+\min }$. Finally, we discussed the stability of 4-dimensional
solutions under dynamical perturbations in the case of asymptotic AdS
spacetime where $z=1$.

In this work, we focused on static asymptotic Lifshitz topological black
holes in the context of Einstein dilaton gravity in the presence of
electromagnetic fields. This work can be extended from different aspects.
First, one may be interested in studying the effects of rotation on
thermodynamics and thermal stability of the solutions. Second, it is
interesting to investigate the thermodynamics of nonlinearly charged
topological Lifshitz black holes with hyperscaling violation. Also, one may
generalize these kind of solutions to other kind of nonlinear
electromagnetic fields. Furthermore, this work can be done in higher
curvature gravities. Some of these works are under investigation.

\acknowledgments{We thank the referee for constructive comments which helped us improve the paper significantly.
We also thank the Research Council of Shiraz University. This work has been financially supported by the Research Institute for
Astronomy \& Astrophysics of Maragha (RIAAM), Iran.}

\end{document}